# A Security Framework for SOA Applications in Mobile Environment

Johnneth Fonseca, Zair Abdelouahab, Denivaldo Lopes and Sofiane Labidi

Federal University of Maranhão, CCET/DEEE
Av. Dos portugueses, Campus do Bacanga, São Luis – MA 65080-040
*johnneth.sfonseca@gmail.com, zair@dee.ufma.br, denivaldo.lopes@gmail.com, labidi@uol.com.br*

## ABSTRACT

*A Rapid evolution of mobile technologies has led to the development of more sophisticated mobile devices with better storage, processing and transmission power. These factors enable support to many types of application but also give rise to a necessity to find a model of service development. Actually, SOA (Service Oriented Architecture) is a good option to support application development. This paper presents a framework that allows the development of SOA based application in mobile environment. The objective of the framework is to give developers with tools for provision of services in this environment with the necessary security characteristics.*

## 1. INTRODUCTION

Over the last years there is a great improvement of capabilities of mobile device, both in its storage capacity and in processing power. This has enabled a wider acceptance of these devices which now offer a variety of applications to users. In addition, new communication technologies allow these devices to access the Internet more efficiently and to communicate with each other.

Actually, it is possible to develop and install in these equipments other applications and services beyond those already coming from the factory. These types of applications may act as service consumers or/and service providers. Thus, there is a need to use a pattern of development that allows developers to create and provide its services more quickly and efficiently. Service-Oriented Architecture (SOA) has emerged as a solution to this type of necessity [13][14][17].

The aim of this paper is to describe a framework to the development of SOA based applications in mobile environment drawing the complexity of their development, with mechanisms to perform all necessary functions for provision of services, such as describing services, carry messages from the parser with specific format, creating a channel of communication to receive and send messages. With this framework, services may be associated with security properties such as cryptography, digital signatures.

This paper is structured of the following form. The first section presents the aim and motivations of the work. The second section describes the SOA architecture and its main components. The third section describes the necessary services in mobile environment. The fourth section shows the mains problems related to security in the mobile environment. The fifth section describes the proposed architecture and the sixth section shows some work related to the proposed work. Finally the last section presents the conclusions and suggestions for future work.

## 2. Service Oriented Architecture (SOA)

SOA describes the keys concepts of software architecture and their relations, where a service and its use are the key concepts that are involved, following a model of publishing services and applications and their universal access [5][13]. SOA has an interface that describes a collection of operations accessible over the network via a standardized format (e.g. XML). These requirements are activated anywhere in a dynamic computing environment and/or pervasive computing where service providers offer a range of services.

SOA creates an environment in which distributed applications and components may create independently of language and platform and focuses on the use of a relatively widespread pattern of communication between operations, enabling thus a model for homogeneous distribution and composition of components.

SOA is a model of components, providing an environment for building distributed systems [6]. SOA applications communicate functionally as a service to the end user's applications and other services, bringing the benefits of low coupling and encapsulation for the integration of enterprises applications. SOA defines the rules of the participants as provider of services, customer of services and registry of services. SOA is not a rating and many new technologies such as CORBA and DCOM at least already had this idea. Web services are new to developers and are the best way to achieve and develop an SOA.

### 2.1. SOA Architecture

The basic architecture of SOA consists of three main components [3] (figure 1):

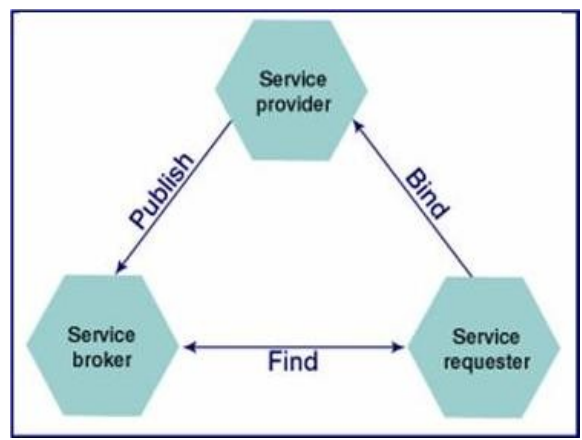

Figure 1.  Basic architecture of a SOA

- Service Requestor (Client) – this entity requires certain functions to perform some task, application or service that relies on interaction with a boot or some service;

- Service Provider – this entity creates and provides the service, it also makes a description of the service and publishes it in a central registry;

- Service Registry (Broker) - location of service description (i.e. where the Service Provider  publishes  a description of the service).

Although a service provider is implemented in a mobile device, the standard WSDL can be used to describe the service, and the standard UDDI registry may be used to publish and make the service available. A challenge is in developing mobile terminal architectures such one of a standard desktop system, taking into account low resources of mobile device [7][15].

## 2.2. SOA Operations

The components of SOA interact with each other through operations (figure 1) which are described below:

- Publish - records a description of the service in directory services, covering the registration of its capabilities, interface, performance and quality that it offers offers;

- Find – searches for services registered in directory services, provided they meet the desired criteria and it can be used in a process of business, taking into account the description of the published service;

- Bind - this operation relies on the service requested or boots an interaction with the service at runtime using the information obtained in discovery of the service;

## 3. WEB SERVICE PROVISION – MOBILE HOST

Mobile Host is a provider of services (***Light Weight***) built to run on mobile devices such as smart-phones and PDAs [1], developed as a *Web Service Handler* built on top of a normal Web server. Mobile Host opens a new set of applications yet little explored [2]. They may be used in areas such as location-based services, community support for mobile and games. It also allows smaller mobile operators increase their business without resorting to a stationary infrastructure. However, these additional flexibilities generate a large number of interesting questions for surveys which require further investigation. Figure 2 shows the main components of a Mobile Host.

The design of a "Mobile Host" is going through many things, some issues where there is very little research; so far set up service provisioning is very limited to devices. The work in [4] describes a model for the development of a Mobile Host system in general.

Traditionally, mobile systems have been designed as client-server systems where **thin clients** such as PDAs or phones are able to use wireless connections to gain access to resources (data and services) provided by central servers [2]. With the emergence of wireless networks, Ad-Hoc and powerful mobile devices it becomes possible to design mobile system using an architecture peer-to-peer [16][18][19].

According to [2], the following characteristics must be guaranteed so that SOA can be built in the mobile environment:

1. The interface must be compatible with the interface of SOA used in the desktop environment for customers;

2. The space used by the service should be small in relation to the mobile device;

3. The service should not affect normal operations of the device;

4. A standard Web server that handle requests of network;

5.  A provider of basic services for treatment of requests for SOA;

6.  Ability to deal with competing requests;

7.  Support the deployment of services at runtime;

8.  Support for the analysis of performance;

9.  Access the local file system, or any external device like a GPS receiver, using infrared, Bluetooth etc.

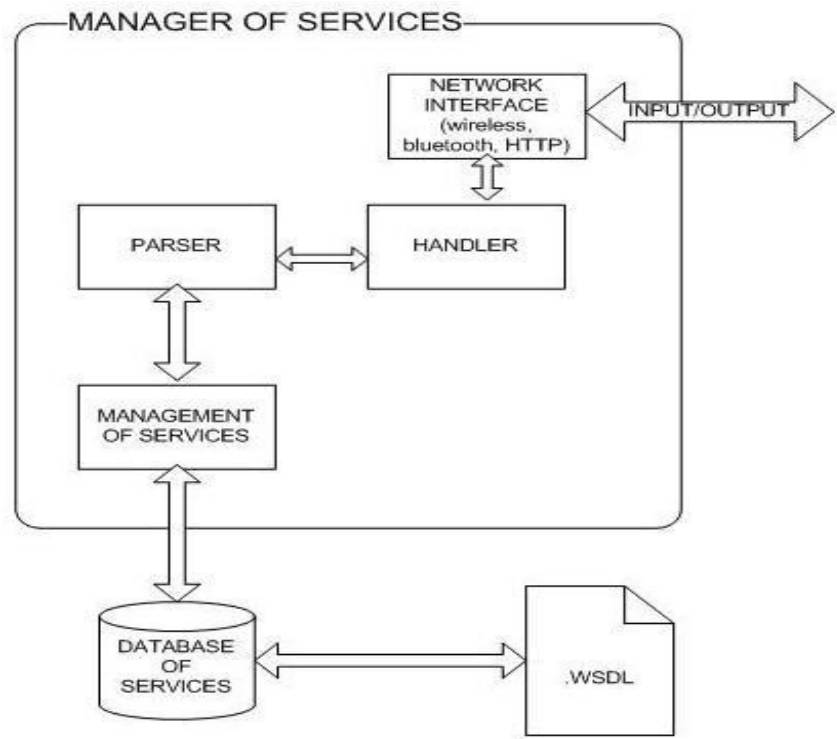

Figure 2.  Core of a Mobile Host

## 4. SECURITY

Security in wireless networks always is evolving. With adequate time a persistent cracker is capable of invading a wireless system. Moreover, some attitudes need to be taken to hinder as much as possible the work of an intruder, allowing basic services of security are met.

Risks already common in wired networks are incorporated into the wireless networks, new arise due to differences in physical structure of these and how they operate. Thus, any solution targeted for wireless networks are to be built in compliance with these new risks because that they are unique to wireless networks.

The greatest threat to a mobile network is the possibility of installing wires through doors in phone calls and data traffic. This threat can be remedied in part with the use of encryption. Consequently, the probability of threat depends on the strength of the encryption algorithm. This resistance is an exit that becomes questionable in the GSM system. Another critical threat, although more hypothetical, is amending the original mobile traffic. In this case the attacker overwrites the data with their own information.

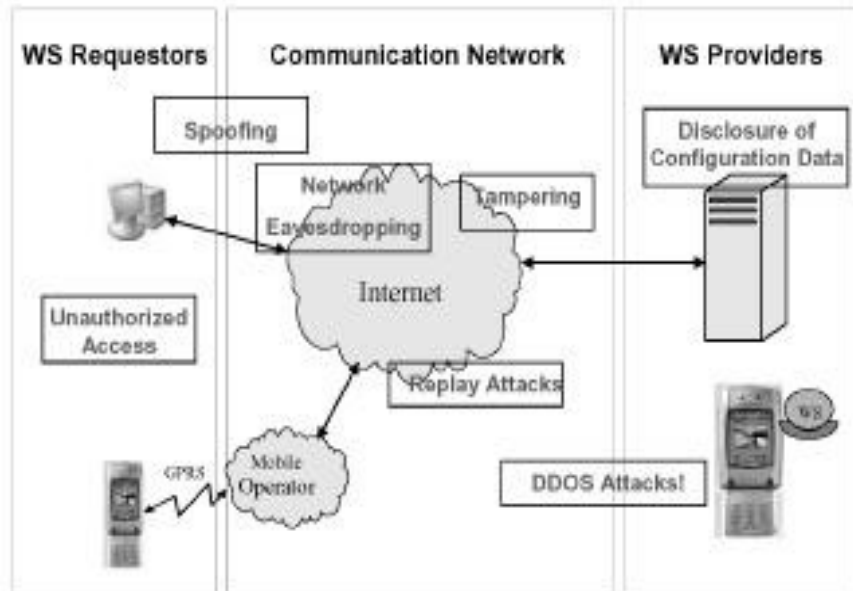

Figure 3 – Typical breaches of security in SOA Mobile

The monitoring of traffic between the device and base station can get the position, speed, duration of traffic, duration, identification of a mobile device. However, the scenarios of exploitation by intruders are the greatest benefit from limited information can be possibly details of location and profile of the user.

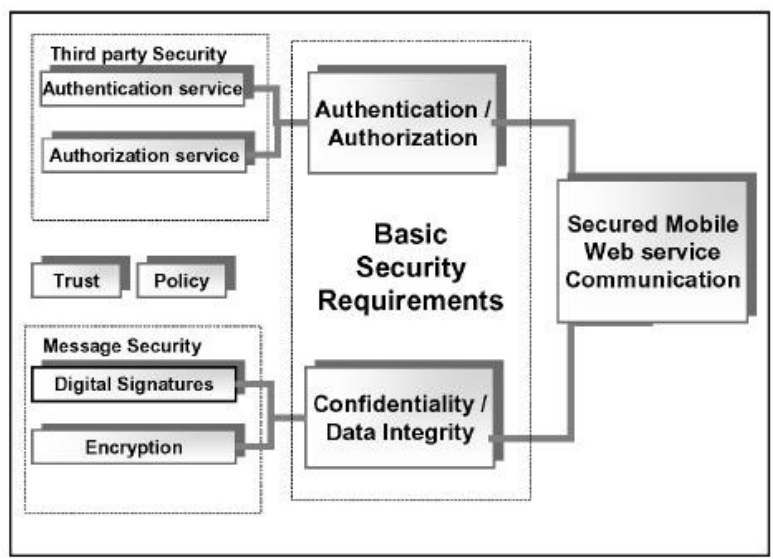

Figure 4.  Basic requirements for the safety of Mobile SOA

Since a SOA is implemented as a Mobile Host, the services are prone to different types of security breaches: such as denial of service attacks, man-in-the-middle, and spoofing of intrusion, and so on. SOA in mobile environment using technologies based on message (such as SOAP over HTTP) for complex operations in several areas. Also, there may be many legitimate services intermediaries in the communication between doing composes a particular service, which makes the context of a security requirement end-to-end.

The need for sophisticated message-level security *end-to-end* becomes a priority for a mobile web service. Figure 3 illustrates some of the typical violations of security in SOA environments in wireless [12].

Considering the breaches of security, the SOA mobile communication must contain at least the basic requirements of security, as shown in Figure 4. Secure transmission of messages is achieved by ensuring the confidentiality and integrity of the data, while the authentication and authorization will ensure that the service is accessed only by trusted requesting. After the success of the implementation of such basic requirements of security, confidence and politicies may be considered as services for mobile field. Political trust can ensure a correct choreography of services. It sets any general policy statements on security policy of insurance SOA, while building relationships of trust in SOA allows safe for the exchange of security keys, providing an appropriate standard of safety.

## 5. OVERVIEW OF ARCHITECTURE

The proposed solution is the development of a framework based on a Mobile Host. A developer can implement the proposed framework and thus make their services (with multiple services in one device) available to the general public without the need of additional deployments to the activities necessary for the provisioning of service, such as receive and send messages, conduct parser of messages to/from the format SOAP/XML, publication of services, generation of WSDL containing information of public services, creation and tracking of an interface for communication to transmit messages on different types of existing technologies (Bluetooth, wireless, HTTP), to identify and implement a requested service. Furthermore, the architecture provides a model of communication secure end-to-end based on public keys. Figure 5 shows the mains components of the proposed architecture, which will be described below:

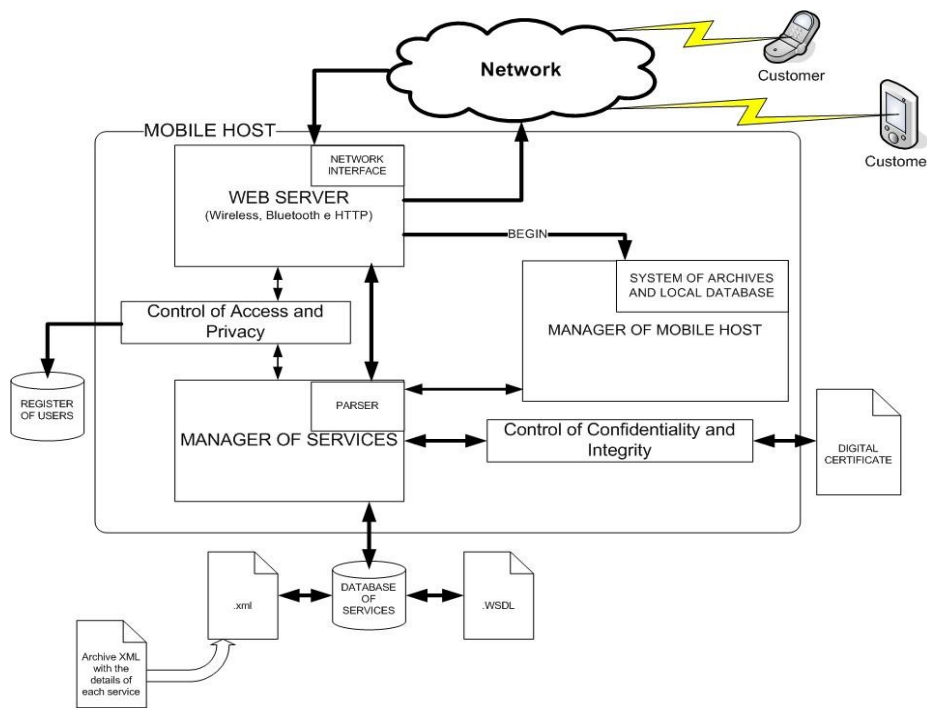

Figure 5.  Framework Proposed

- **Web Server** - provides the functionality of a web server to the mobile device, through the framework of this component receives and sends messages according to the type of technology being used (Bluetooth, wireless or HTTP). Also, check if the message received is a SOAP message or a normal web request, passing it on Manager Services if a SOAP request or in the case the request is a normal web;

- **Control of Access and Privacy** - responsible for verifying whether a consumer can access the services available on the device;

- **Manager of Services** - responsible to provide the functionality necessary for the provision of services in the mobile device. This component performs activities such as the description of the service after reaching the details of it and the generation of a WSDL document related thereto; storage of the service in a local database, called Database Services; generation of document for its publication. Also receiving the SOAP requests and processes, checking possible errors, obtaining and executing the requested service;

- **Manager of Mobile Host** - responsible for setup and initiation of the other components of middleware. In this module the databases used are created and available for use with other components, verify that a service has been created, or not, and send data from a new service to be processed by the Manager of Services;

- **Control of Confidentiality e Integrity** - adds mechanisms that allow messages sent by consumers remain confidential, i.e., can't be interpreted by others through the use of pairs of symmetric keys. In addition, this module allows the framework to generate a digital signature created for each message, which allows the broadcast content is changed if it becomes possible to identify it and also of digital certificates means by which the consumer gets the key issues of service;

- **Databases**: The framework uses some database to store certain information that is necessary for carrying out their activities. These database, and their features are listed below:

- **Database of Services** - This base is used to store the services created and are available in the device;

- **Database of Users** - responsible for storing the data for validation of consumers made through the registration the login and password hash. The data for identifying a device and if they have permission to access the services also are stored on this database;

The proposed architecture can be structured into a model composed of the following layers (see figure 6):

1. Network Layer Interface - responsible for creating a channel of communication between the provider and consumer service, receiving and transmitting messages and can use different types of available communication technologies;

2. Events Layer - verify what kind of message is being transmitted and therefore can perform activities related to it, is also responsible for conducting the parser of messages to/from the format SOAP/XML;

3. Service Layer - through this layer a service can be described, made available for publication, obtain and enforce a method when it is invoked;

4. storage Layer – place where all created services and the latest user requests are stored; this layer stores as well WSDL descriptions and the document object containing the information of each created department;

5. Security Layer - ensures that there is security end-to-end in every step since the transmission of a request and the response of the request is made by the end user;

6. Management Layer - allows the management of all services offered by the developed framework;

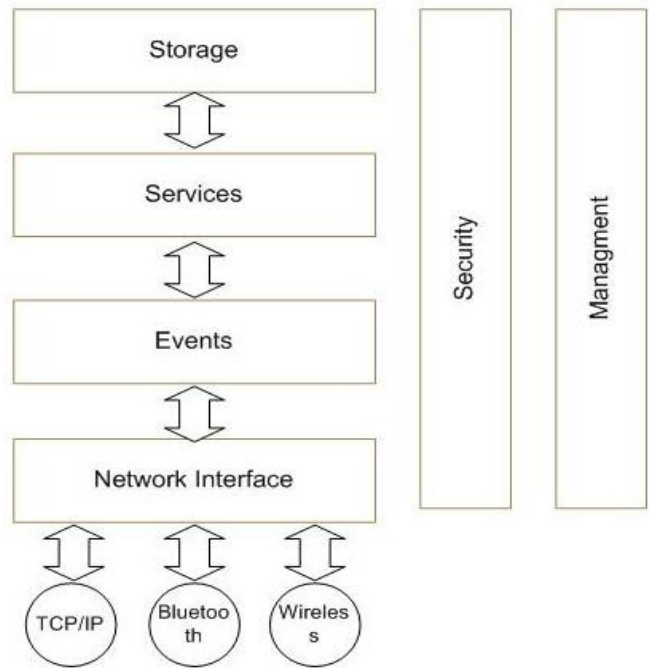

Figure 6.  Model in Layers

## 5.1. Process Development Services

A service is created by the framework must implement the interface *iService* to describe the service, and thus informing what methods, their parameters and return type. In addition, A user must use the method *executeMethod* to inform how the methods will be executed, because *executeMethod* is called by the system to run method after going through all stages of verification. In this method it is created an instance of object where the service is implemented or of the class that implements *iService*.

After the developer has informed the details of the service, the system automatically generates a WSDL document, which contains the details of the service following a pattern already known, and stores in a specific folder on the device itself for further verification, or can be sent to a consumer. Figure 7 shows the mapping of a Java class to a WSDL document that contains a description.

In case the developer wants to add security services the developer must instantiate to class "*SymmetricKey*" responsible for the generation of a pair of symmetric keys (public and private) used to create a digital certificate (which can be sent to Final consumers) for the service and create a digital signature for each message sent, to guarantees its integrity.

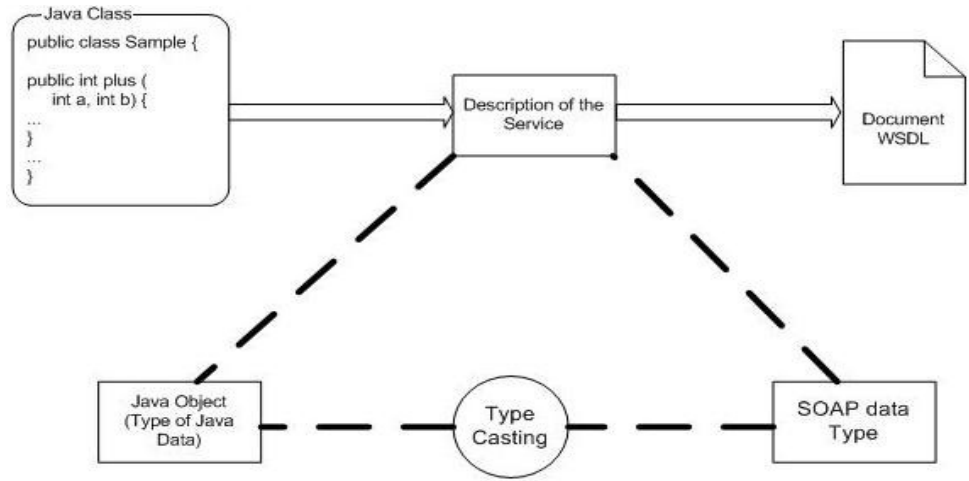

Figure 7. Mapping Java class for WSDL document

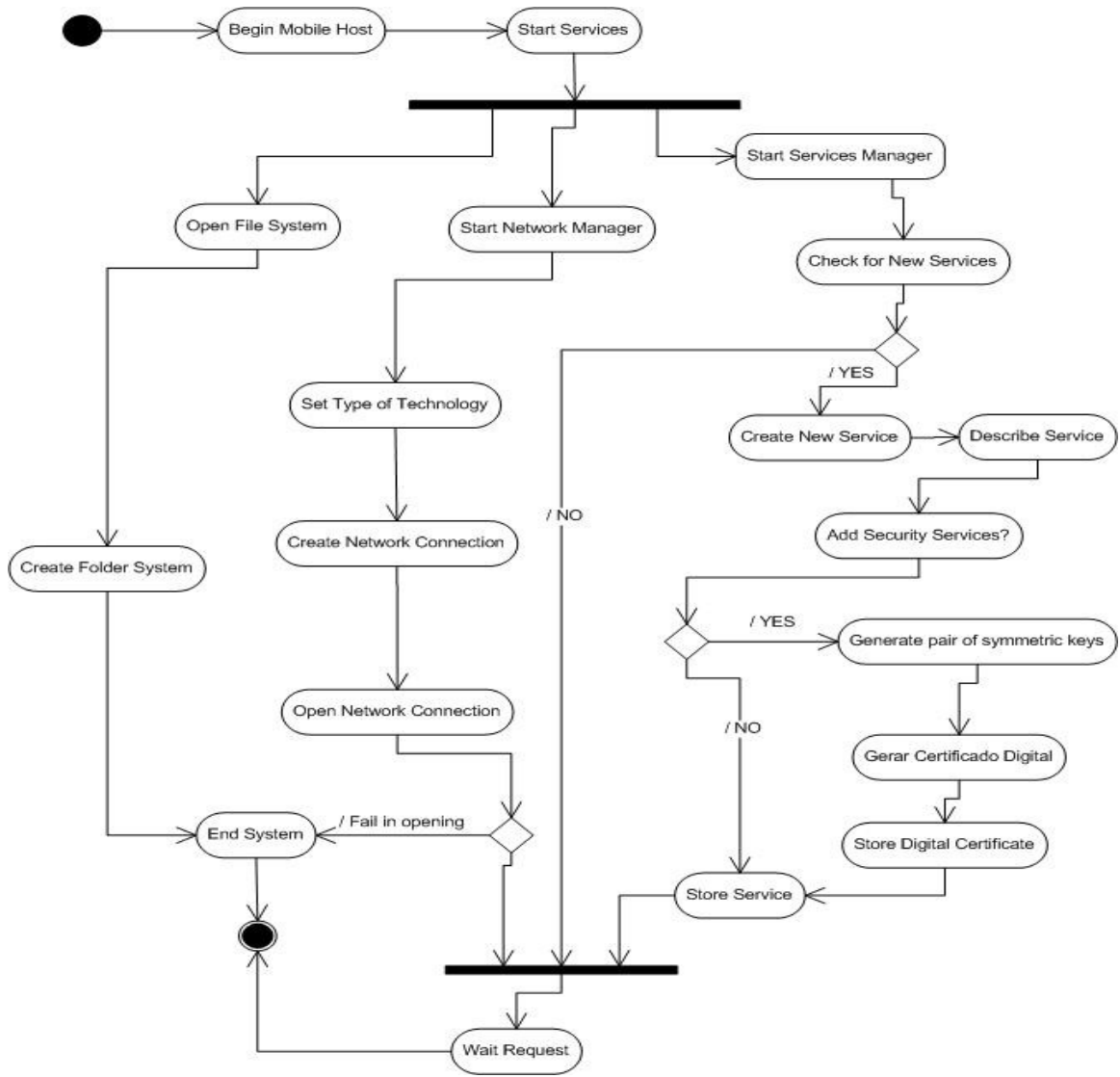

Figure 8. Activity Diagram – Mobile Host -- Creation of Services

Following the steps outlined above the service will be stored in a repository specifically for its further use. Several services can be created and stored, allowing the developer holds a range of services to make them available to their consumers, but the restrictions are limited memory capacity of the device and where it will be allocated. Figure 8 shows a diagram of activities with for the startup process of a Mobile Host and creating new services.

## 5.2. Network Interface

The framework guarantees the communication between the service provider and the end consumer because the framework has a mechanism that provides this features. This mechanism allows different technologies can be used for communication (Bluetooth, HTTP or Sockets). The developer can choose which fits that their needs and provide the network address and port that will be used, or can use the default values set by the framework.

Once started, the Network Interface Service stay listening to demands on the network and it is responsible for receiving and sending messages to/from the Mobile Host, and thus behaving a standard Web server. Each received request a new line of enforcement is designed to meet the request to identify if it is a SOAP request or a normal web request.

## 5.3. Invocation Service

When the message received is a normal web request the web server treats it normally.

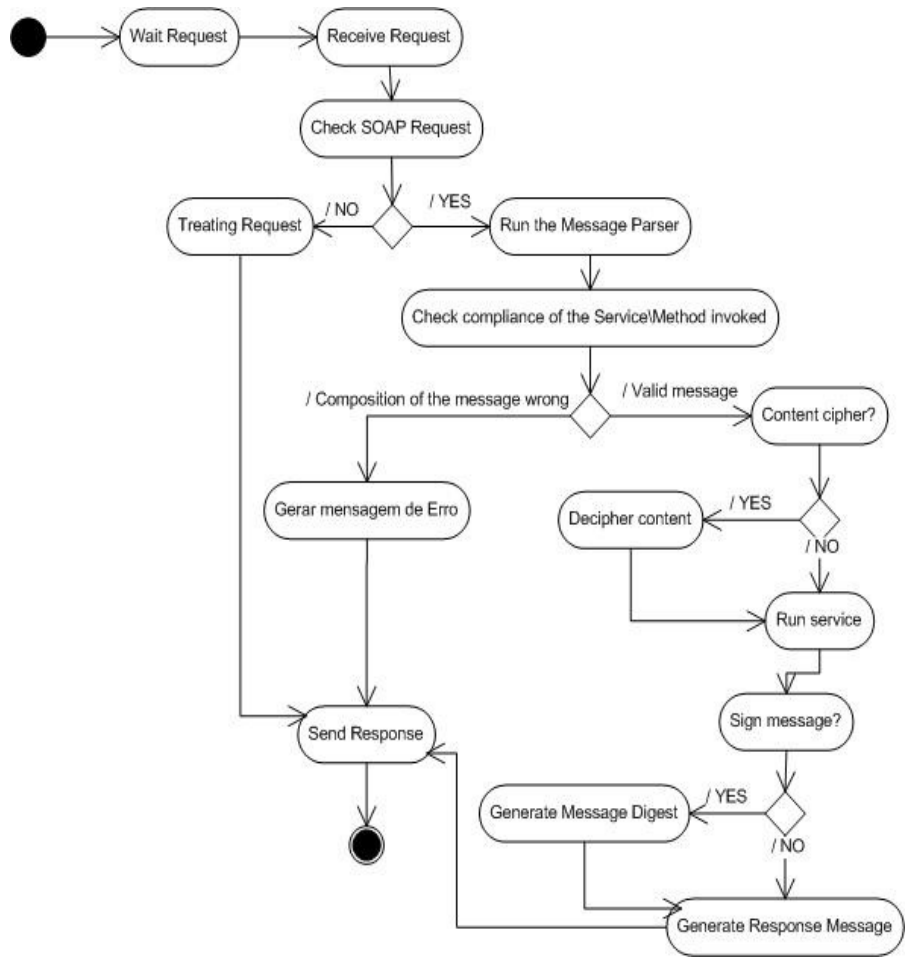

Figure 9.  Activity Diagram – Service Invocation

However if the request is a SOAP request its treatment is passed to a handler for the SOAP requests, which performs the message parsing, extracting the information necessary for the invocation of a service. During the process of parsing, it is checked if the invoked service is compound all the requirements of the method invoked (quantity and types of variables, besides the type of return, according to the SOAP specification) according to the data reported by the developer in the process of creating the service.

If the message is not compatible with the requirements, a error message is generated by the system and send it to the customer, otherwise the method is executed through the invocation of the ***executeMethod*** to which are passed the method name and parameters from the message. Once in possession of the response of the implementation of the method invoked a reply message is generated and sends it to the customer. If the developer has added security services, a digital signature is generated for the response of the Mobile Host and sent along the message. Figure 9 shows the diagram of activities with the process of invocation of a service.

## 5.4. Parsing of messages

An important feature of the Mobile Host is the ability to perform the parsing of messages received as a SOAP envelope to an object that contains the details of the request and that is used to obtain the information necessary to implement the request. Also, the reverse process is possible, that is the response is converted into a document SOAP before being sent to the requester. The framework also generates error messages (SOAP FAULT) if any failure is identified in the request.

## 5.5. Security Service

The model described in this paper proposes a security mechanism based on the content of messages, which is engaged in protecting the communications end-to-end between the consumer and service provider, ensuring that messages delivered are not corrupted by third party. This mechanism provides tools for generation of digital signatures and encryption of transmitted messages. The framework can create pair of keys (public and private) for each service provided by the Mobile Host during the process of creating the service. Though, this is optional the final developer decides whether or not this service can be added to its system.

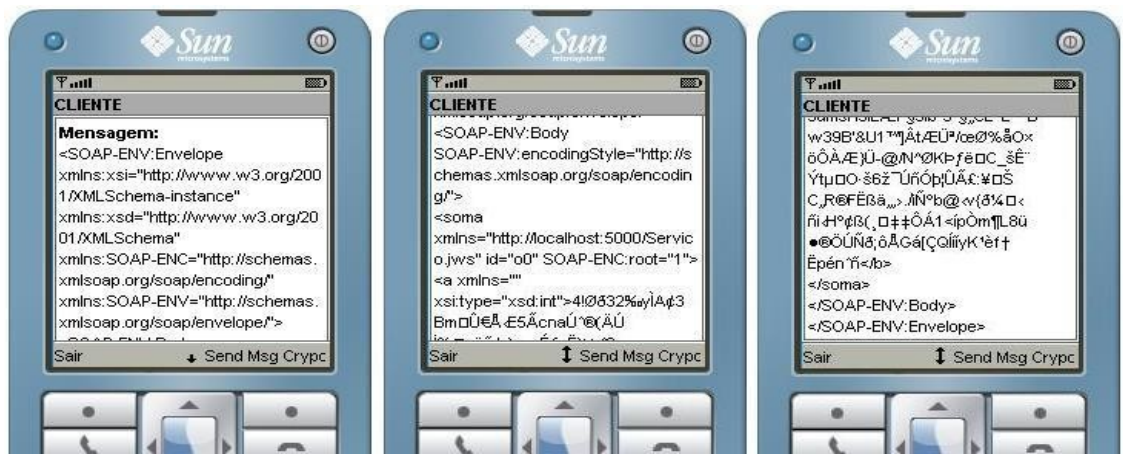

Figure 10. Sample of Encrypted Message in Mobile Device

The encryption based on public key can also be used in the process of digital signature, which ensures the authenticity of who sends the message, associated with the integrity of its contents. Through these keys it can be created a digital signature to be sent along with a message, and also for allowing the transmitted messages to be encrypted and decrypted as observed in Figure 10.

To make the public key available for a possible consumer, the system adds the generation of digital certificate that can be sent to mobile consumers, through which they get the public key of the services which he wants to communicate, and thus can encrypt the messages and verify the authenticity of digital signatures received by them. An example of a Digital Certificate generated by the system can be seen below:

---------- Begin Certificate ----------
Type: X.509v1
Serial number: 35:32:35:38:30:39
SubjectDN: MobileHost/
IssuerDN: MobileHost/
Start Date: Wed Aug 13 17:37:58 UTC 2008

Final Date: Sat Aug 23 17:37:58 UTC 2008
Public Key: RSA
modulus:
9337497433152274214014446837013394548439412
0408072691949673228126709006135379056199872
8561435094844888900388751758 ...
public exponent:65537
Signature Algorithm: RSA
Signature:
2673219510732109272348323117338219317732109
2162271232116242582531891517420121224291208
4101821296251673769631166216239 ...
---------- End  Certificate ----------

## 6. CASE STUDY

This section presents some case studies that were developed with the aim of assessing the implementation of the proposed framework. The services are created in accordance with the roadmap presented in section 5.1.

### 6.1. Note System

This case study presents a service which provides the notes of a group of students in certain disciplines stored on a mobile device through an application installed on the device itself, which can be seen in Figure 11 (this application runs on Mobile Host).

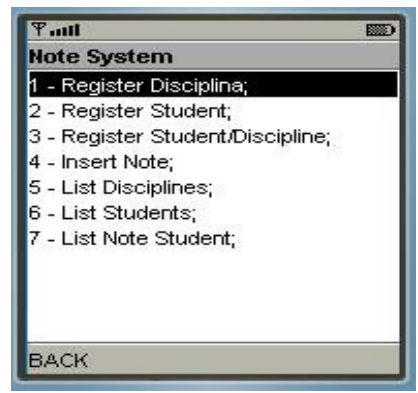

Figure 11.  Application of Service System Notes

When a client wants to obtain the notes of a student he must create a SOAP request and sends it to the device where the service is installed. Figure 12 shows the consumer informing the data which will be used in the request.

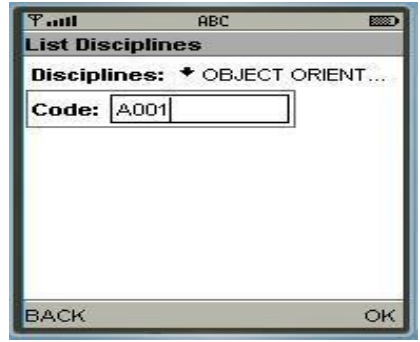

Figure 12.  Consumer service informing the data to create a requisition

```
– <SOAP-ENV:Envelope xmlns:xsi="http://www.w3.org/2001/XMLSchema-
    instance" xmlns:xsd="http://www.w3.org/2001/XMLSchema"
    xmlns:SOAP-ENC="http://schemas.xmlsoap.org/soap/encoding/"
    xmlns:SOAP-ENV="http://schemas.xmlsoap.org/soap/envelope/">
  – <SOAP-ENV:Body SOAP-
      ENV:encodingStyle="http://schemas.xmlsoap.org/soap/encoding/">
    – <obterNotas xmlns="http://localhost:5000/CadastroEscolar.jws"
        id="o0" SOAP-ENC:root="1">
        <codAluno xmlns="" xsi:type="xsd:string">A001</codAluno>
        <codDisciplina xmlns=""
          xsi:type="xsd:string">D002</codDisciplina>
      </obterNotas>
    </SOAP-ENV:Body>
  </SOAP-ENV:Envelope>
```

Figure 13.  SOAP message with the request of Notes

With this information a SOAP request may be generated as we see in Figure 13. Thus, the message is sent to the Mobile Host, where the service is installed. When this message arrives in the Mobile Host and it is identified as a SOAP request, it is then forwarded to the service manager which is responsible for the treatment. Initially, a parser is invoked to obtain the requested service and extract the data. An analysis is done to check whether the service exists and the data is conforming to the existing service.  If it is the case, the service is performed and the results are returned through a message which is generated and is sent back to the requester (Figure 14). If the service does not exist or the data is incorrect or a method failure occurs during the execution of a service, an error message is generated and sent back to the requester.

When the response is received, the application service consumer extracts its content and displays the information obtained to the user's device, as can be seen in Figure 15. The application must also be prepared to receive and display an error message, if it occurs during any treatment of any request.

```
<?xml version="1.0" encoding="utf-8" ?>
- <soap:Envelope xmlns:xsi="http://www.w3.org/2001/XMLSchema-
    instance" xmlns:xsd="http://www.w3.org/2001/XMLSchema"
    xmlns:soap="http://schemas.xmlsoap.org/soap/envelope/">
  - <soap:Body>
    - <obterNotasResponse xmlns="http://www.dee.ufma.br/">
        <obterNotasResult
          xsi:type="xsd:string">#A001;D002;LACKS:;0#A001;D002;FINAL
          TEST:;0#A001;D002;REPLACEMENT:;0#A001;D002;NOTE
          3:;98#A001;D002;NOTE 2:;95#A001;D002;NOTE
          1:;100#</obterNotasResult>
      </obterNotasResponse>
    </soap:Body>
  </soap:Envelope>
```

Figura 14. Message response

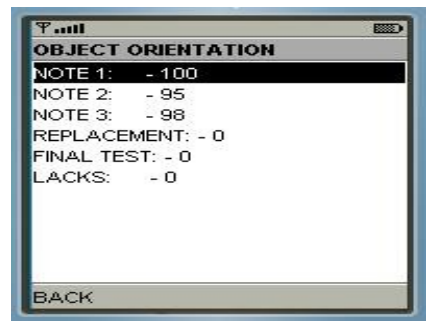

Figure 15 – Information displayed to the user's device (Response)

# 7. RELATED WORKS

The idea of provisioning SOA for mobile devices is explored in [4]. According with [4], the basic architecture of a mobile terminal as a service provider can be established through a: Service Request, Service Provider (Mobile Host) and service registry. The service providers implemented in the mobile device are used to describe the standard service (WSDL), and a standard (UDDI) is used to publish and remove publication services. Once you can see that this architecture follows the pattern of desktop systems, but taking into account the low resources of the device.

According with [4] with the advent of wireless and ad-hoc networks, there is increase in capabilities of equipment that allow the creation of entirely pure P2P networks, where the key piece of this architecture is to maintain the Mobile Host completely compatible with the interfaces used by (WS) SOA in such a way customers do not perceive the difference; developing Mobile Host with few resources available on the device, and limit the performance of functions of SOA in such a way it does not interfere with the functions of the device.

The core of the architecture proposed by [4] is: a *network interface which* is responsible for receiving the requests and sends the answers to consumers, *Request Handler* extracts the contents of the request and sent them to the *Service handler* that accesses the database of services and executes the requests.

In [9], it is proposed a mechanism for hierarchical service overlay, based on the capabilities of the hosting devices. The Middleware of [9] is called PEAK (Pervasive Information Communities Organization) which enables the transformation of resources into services. Based on the widespread hierarchical "overlay service" by developing a mechanism for composition of

104

services which is able to make dynamically complex services using the availability of basic services.

In [10], it is developed a mobile middleware technology, motivated primarily by projects related to mobile applications. Using this Middleware, mobile devices become a part of SOA (Web Services) located on the Internet. The Middleware enables not only access to SOAs on the Internet, but also the provisioning of services. These services can be used by other applications of the Internet (mobile call back services) or any other fixed or mobile device (P2P services).

In [11], the author investigates mechanisms to support dynamic m-services oriented architecture context with the objective of service publication and dynamic discovery of mobile users anywhere and anytime discussing the characteristics of SOAs for wireless and wired networks. Moreover, [11] investigates the availability of technologies for mobile device and SOA paradigm; proposing an m-services architecture to support dynamic and scalable mobile services. [11] proposes an entity management services with a service record as an intermediate layer between service providers and mobile users. The management service is responsible for coordination of interactions between service providers and mobile users. This interaction produces information services, providing and delivering services for mobile users at any time. [11] is also investigating the use of dynamic invocation interfaces as a mechanism of communication between departments of the description and invocation at runtime.

## 8. Final Considerations

This paper proposes the use of devices as service providers; however its development requires an implementation work that can become extremely complex. Therefore, we see the need for a tool for development and provision of services on mobile devices in an automated way.

This work introduces the Mobile Host as part of a Middleware for construction of SOA in mobile environment. This paper presents an overview of the concepts, and proposes an architecture. The architecture proposal aims to provide a developer with a tool for rapid development of services in mobile environment and aggregates all services necessary for the provision of the service for invocations in an infra-structured network such as in P2P networks.

A security service is included in the framework, where the developer has the option of whether or not to use it. While it is a very important requirement, its use requires a greater use in processing power of the device and the response time of a request. Then it should be considered whether its use is feasible for each service being developed. The framework presented proved to be a good solution to the problem presented. However, more tests related to their performance and quality of service offered yet become necessary.

### Acknowledgements

Financial Support of CNPq and Fapema are gratefully acknowledged.

## AUTHORS

**Johnneth Fonseca** has obtained her B.Sc. degree in Computer Science from UFMA, Brazil in 2006 and hir M.Sc. degree in Computer Science from UFMA, Brazil in 2009. His research interests include networking, and security.

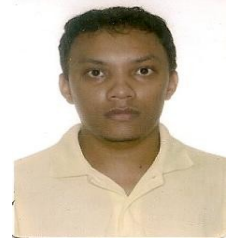

**Zair Abdelouahab** is a professor of Computer Science at the Federal University of Maranhão (UFMA) in Brazil. He is conducting research in Distributed Systems, Networking, Security, Requirement and Software Engineering. He received his BSc degree in Computer Engineering from University of Setif, Algeria in 1985, his M.Sc. degree in Computer Science from Glasgow University, UK in 1988 and a Ph.D. degree in Computer Science from Leeds University, UK in 1993.

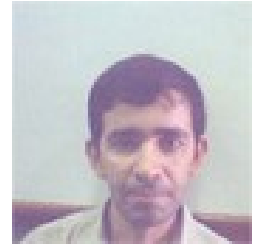

**Denivaldo Lopes** is a professor at the Federal University of Maranhão (UFMA) in Brazil. He received his B.Sc. degree in Electrical Engineering from UFMA in 1999. He received his M.Sc. in Electrical Engineering, sub area Computer Science, from UFMA in 2001. He received his Ph.D. degree in Computer Science from the University of Nantes/France in 2005. His research interests include Service-Oriented Architecture, Model Driven Engineering, Security and Telemedicine.

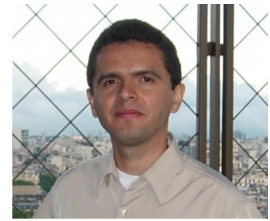

**Sofiane Labidi** is a professor at the Federal University of Maranhão (UFMA) in Brazil. He received his B.Sc. degree in Computer Science in 1990, his M.Sc. in Computer Science from University of Nice - Sophia Antipolis (France) in 1991 and his Ph.D. degree in Computer Science from INRIA (the French National Institute for Research in Computer Science and Control) and the University of Nice - Sophia Antipolis (France) in 1995. His research interests include Artificial Intelligence, Computer Education and Multiagent Systems.

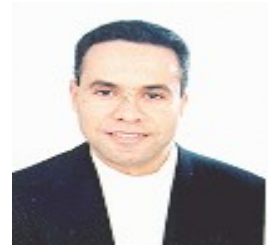

.